\documentclass[ reprint, amsmath,amssymb, aps,prb,]{revtex4-1}

\usepackage{graphicx}
\usepackage{dcolumn}
\usepackage{bm}
\usepackage{color}

\begin{document}

\preprint{APS/123-QED}

\setlength{\abovecaptionskip}{-0pt}

\title{Field Angle Tuned Metamagnetism and Lifschitz Transitions in UPt$_3$}

\author{B.S. Shivaram$^1$, Ludwig Holleis$^1$, V.W. Ulrich$^1$, John Singleton$^2$ and Marcelo Jaime$^2$}

\affiliation{$^1$Department of Physics, University of Virginia, Charlottesville, VA. 22904}
\affiliation{$^2$National High Magnetic Field Laboratory, Los Alamos National Labs, Los Alamos, New Mexico}

\date{\today}

\begin{abstract}
Strongly correlated electronic systems can harbor a rich variety of quantum spin states.  Understanding and controlling such spin states in quantum materials is of great current interest.  Focusing on the simple binary system UPt$_3$ with ultrasound (US) as a probe we identify clear signatures in field sweeps demarkating new high field spin states.  Magnetostriction (MS) measurements performed up to 65 T also show signatures at the same fields confirming these state transitions.  At the very lowest temperatures ($<$200 mK) we also observe magneto-acoustic quantum oscillations which for $\theta=90^0$  and vicinity abruptly become very strong in the 24.8-30 T range.  High resolution magnetization measurements for this same angle reveal a continuous variation of the magnetization implying the subtle nature of the implied transitions.  With B rotated away from the c-axis, the US signatures occur at nearly the same field.  These state transitions merge with the separate sequence of the well known metamagnetic transition which commences at 20 T for $\theta=0^0$ but moves to higher fields as 1/cos$\theta$.  This merge, suggesting a tricritical behavior, occurs at $\theta \approx51^0$ from the ab-plane.  This is an unique off-symmetry angle where the length change is precisely zero due to the anisotropic nature of MS in UPt$_3$ for all magnetic field values. 
\end{abstract}

\maketitle

\textbf{INTRODUCTION:} In an itinerant metamagnet, such as a heavy fermion system, a magnetic field can cause a rapid increase in the magnetization at a critical value, $\rm B_ c$. At sufficiently low temperatures, this may appear as a step change, and thus maybe termed a quantum phase transition (QPT)  \cite{SiScience2010, GegenwartNat2008, LohneysenRMP2007, GrigeraScience2001}.  Many heavy-fermion (HF) metamagnets are highly anisotropic~\cite{AokiCRP2013} with both $XY$-type~\cite{KitagawaPRL2011} ({\it e.g.,} CeFePO) and Ising-type~\cite{HaenJMMM1987, SugiyamaPhysicaB1993} ({\it e.g.,} $ \rm CeRu_2 Si_2$ and $ \rm URu_2 Si_2$) possible.  In such metamagnets a critical field $\rm B_ c$ is observed for a specific orientation, but only a gradual increase being present when the field is in the perpendicular direction. The latter is invariably the hard axis, exhibiting a smaller low-temperature susceptibility, in many cases by nearly an order of magnitude.  $ \rm UPt_3$ with a hexagonal crystal structure is one of the earliest discovered HF metamagnets (MM) and has been studied for more than three decades.  In this MM in addition to the rapid rise in the magnetization at a critical field $ \rm B_c=20 T$ applied in the basal plane ($\theta=0^0$) the longitudinal ultrasound velocity \cite{ShivaramUltrasound2015} suffers an enormous reduction.  Furthermore, the dip in the velocity grows inversely with temperature, saturates below a characteristic low temperature and is asymmetric as revealed in the difference in behavior between $ \rm B<B_c$ and $ \rm B>B_c$. The anisotropic angular variation of $ \rm B_c$ has also been measured \cite{SuslovIntJModPhys2002} in $ \rm UPt_3$.  It follows the same 1/cos$\theta$ dependence found in more `conventional' MMs where the spins are known to be well localized \cite{GoddardJPhys2006}.  However, although $ \rm UPt_3$ is more forgiving, in the sense that the susceptibility along the hard axis is smaller by less than a factor of two compared to its value in the easy plane, it is challenging to measure the full quadrature dependence of $ \rm B_c$ since it rises rapidly to large unattainable values as $\theta$ approaches $ \rm 90^0$.  Thus most measurements such as magnetization with field along the c-axis are featureless and imply a paramagnetic like behavior to the highest measured fields \cite{Franse1990}.  In addition to the enormous changes in the sound velocity at the MM transition in $ \rm UPt_3$, a much weaker signature is seen at 4.5 T and has been attributed to a spin density wave state (SDW) transition \cite{BrulsPhysicaB1996}.  This signature is anisotropic and moves to 9 T for $ \rm \theta=90^0$.  In more recent work we have uncovered a number of exciting new results. The nonlinear susceptibility, $\chi_3$, for $\rm \theta=0^0$ peaks at a temperature $\rm T_3$ which is $\approx 1/2 \rm {T_1}$ the temperature where the linear susceptibility peaks \cite{ShivaramUniversality2014, ShivaramChi52014}.  However, for $ \rm \theta=90^0$  $ \chi_3$ is monotonically negative and featureless.  In the current paper, we present new high field and very low temperature high resolution measurements of the sound velocity, magnetostriction and magnetization in UPt$_3$ which reveal a complex and rich anisotropic behavior.\\

\begin{figure*}
\includegraphics[width=0.9\textwidth]{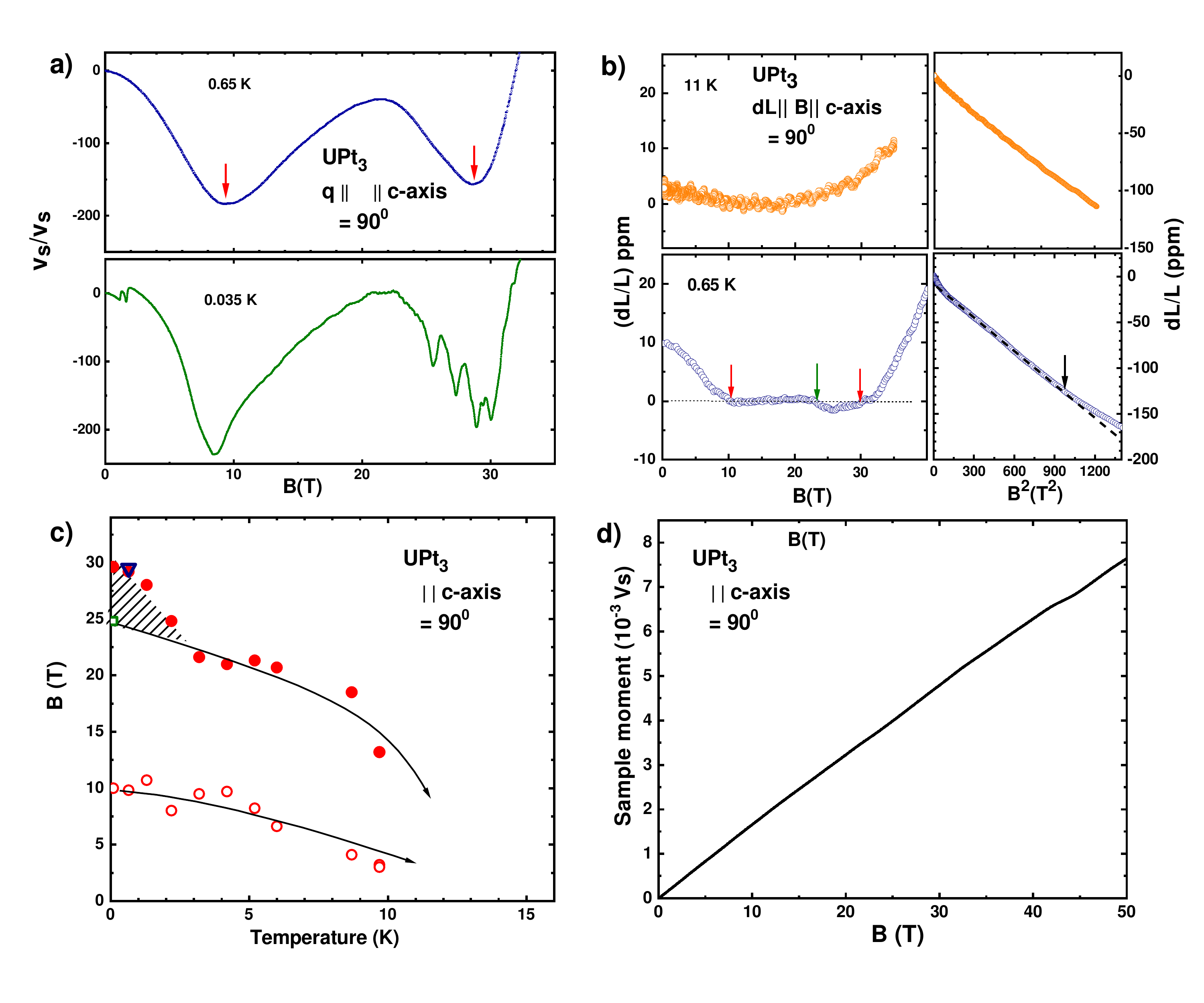}
\caption{\label{fig1}Shown in the top left (panel (a)) are the changes from zero field value in the velocity of longitudinal ultrasound propagating in $ \rm UPt_3$ along the c-axis with B$\parallel$c.  Two dips in the sound velocity at 9 T and 30 T are clearly seen.  Also visible in the 35 mK data are magneto acoustic quantum oscillations.  Panel (b) shows on the right the as measured length changes for B$\parallel$c plotted against $ \rm B^2$.  A clear departure from linearity is seen around B=30 T for the 0.65 K run.  The left side of this panel, for T=0.65 K, shows the same data with the quadratic background removed to highlight features at 9 T and 30 T (red arrows). A smaller feature at 24 T (green arrow) is also apparent. The bottom left panel (c) shows the phase diagram obtained from isothermal velocity scans such as those shown in (a).  The triangle at 30 T is from the magnetostriction measurement.  The two US minima shift to lower magnetic fields as the temperature increases and appear to reach zero on a scale consistent with the temperature where a peak in the sound attenuation has been observed \cite{MullerPRL1986}.  Also note the sudden upturn in the high field state for T$<$ 4K and the hatched sliver in the phase diagram. The open square at 24 T marks the onset of the large MAQO. Although clear signatures of the high field transitions are visible in the ultrasound velocity the magnetization measured (bottom right panel (d)) is continuous across 30 T. }
\end{figure*}

In our study with $\rm UPt_3$ we find that in contrast to an expected featureless response, the `perpendicular' orientation reveals a rich structure in ultrasound velocity which has important implications to the anisotropic evolution of the MM transition.  In high resolution longitudinal ultrasound (US) and magnetostriction (MS) measurements due to their unprecedented sensitivity we uncover minute physical effects missing from either transport, magnetization or other thermodynamic measurements.  The current experiments reveal for the first time a "dip" in the sound velocity with a minimum at 30 T for $\rm \theta=90^0$.  Previous US work for this orientation did not extend to high enough fields and detected only a low field minimum at 9 T \cite{BrulsPhysicaB1996, BrulsPhysica1993}.  We also present magnetostriction and high resolution magnetization measurements performed separately in further characterizing the state transitions implied by the US signatures.  Our MS measurements also extend to fields higher than before and reveal distinct signatures at the same field position as the sound velocity minima at the lowest temperatures. In addition, we present US measurements for fields oriented away from the c-axis from which we obtain the angular dependence of the newly observed transitions.  \\

\begin{figure}
\begin{minipage}[l]{0.4\textwidth}
\hspace{-1,3cm}
\vspace{0.1cm}
\includegraphics[width=80mm]{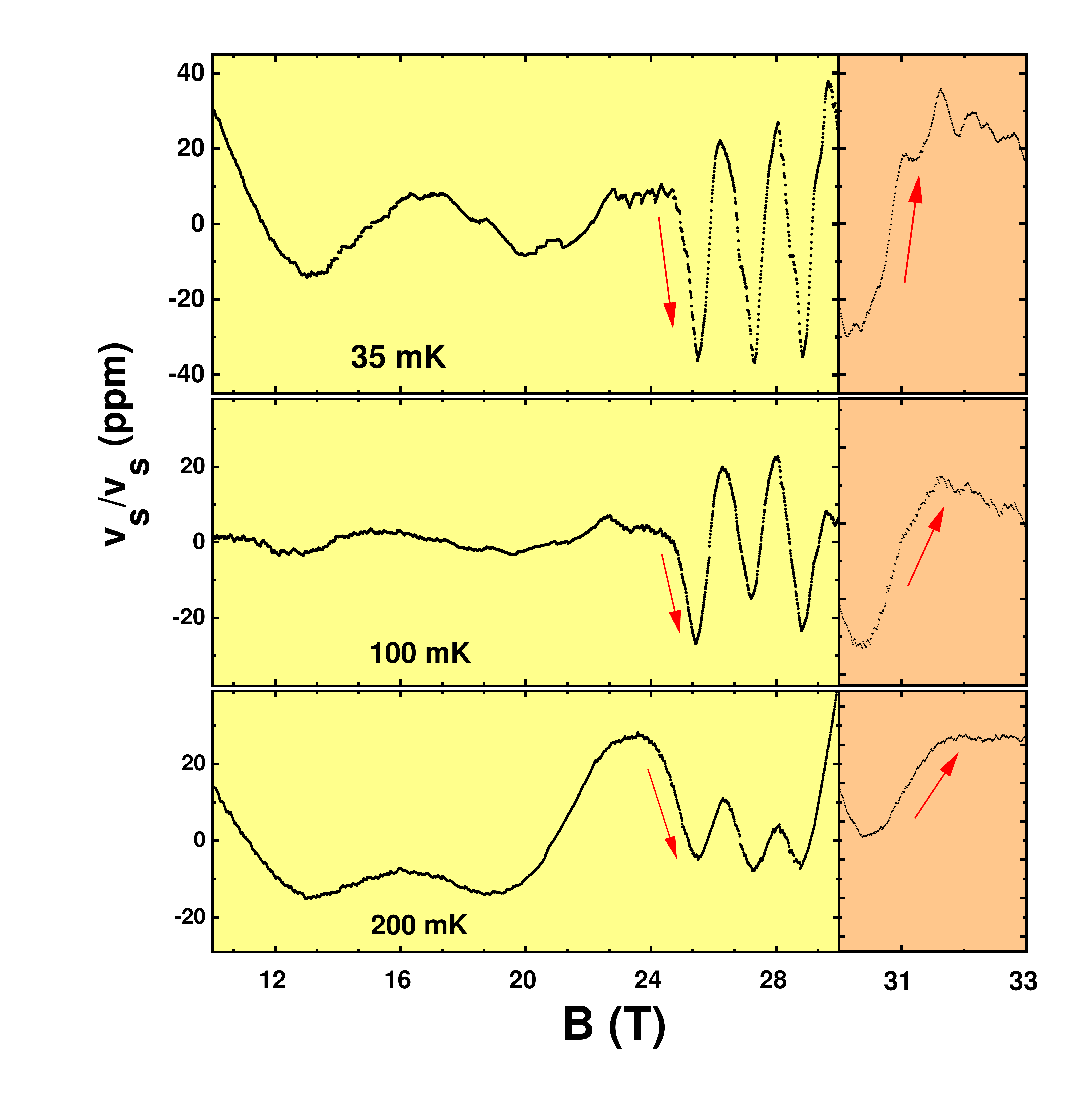}
\end{minipage}
\begin{minipage}[l]{0.4\textwidth}
\vspace{-0.9cm}
\hspace{-0.1cm}
\includegraphics[width=85mm]{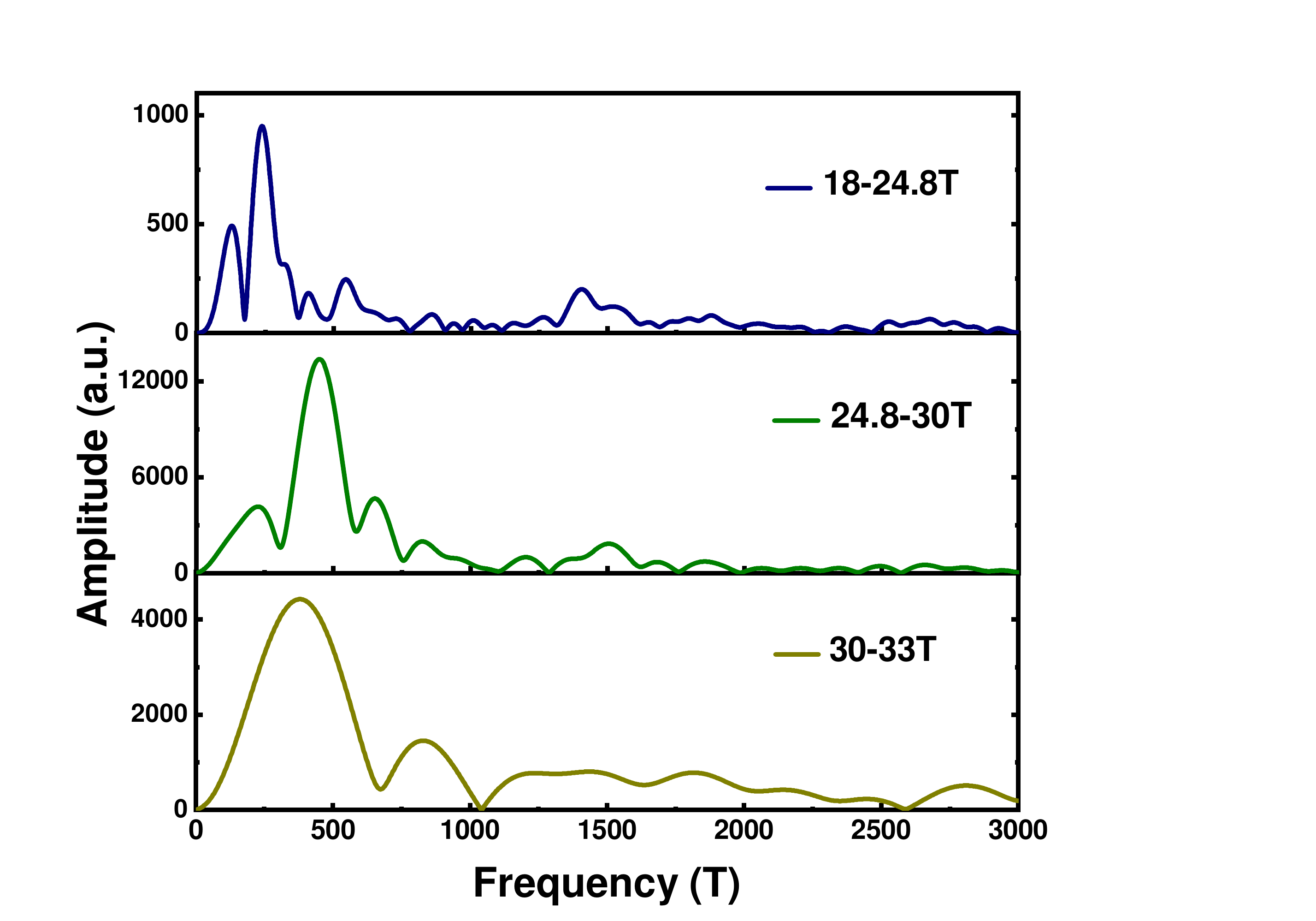}
\end{minipage}
\caption{\label{fig2}The top part of the figure shows the quantum oscillations in the sound velocity, for B$\parallel q\parallel$c-axis at T=35 mK, 100 mK and 200 mK.  To highlight these oscillations we have subtracted a quadratic and a linear background as discussed in supplementary information (fig.S2).  An abrupt increase in the magnitude of the dominant oscillation as well as its period decrease commencing at 24.8 T can be clearly seen. In addition it may be noted that the sharpness of this transition, as indicated by arrows, increases as T$\rightarrow$0 thus establishing it as a possible quantum phase transition.  These facts are further apparent in the Fourier transformation of the lowest temperature data in the three field regions 10-24.8 T, 24.8-30 T and 30-33 T shown in the bottom panel.   Analyzing the rapid decrease in the amplitude of the dominant oscillation ($\approx$ 450 T) with warming yields an effective mass, m$^*$=32 m$_e$.    A clear shift in the frequencies between the adjacent field regions suggests both the transitions at 24.8 T and 30 T are Lifshitz Transitions.}
\end{figure}

\begin{figure}
\begin{minipage}[l]{0.4\textwidth}
\hspace{-1,3cm}
\includegraphics[width=80mm]{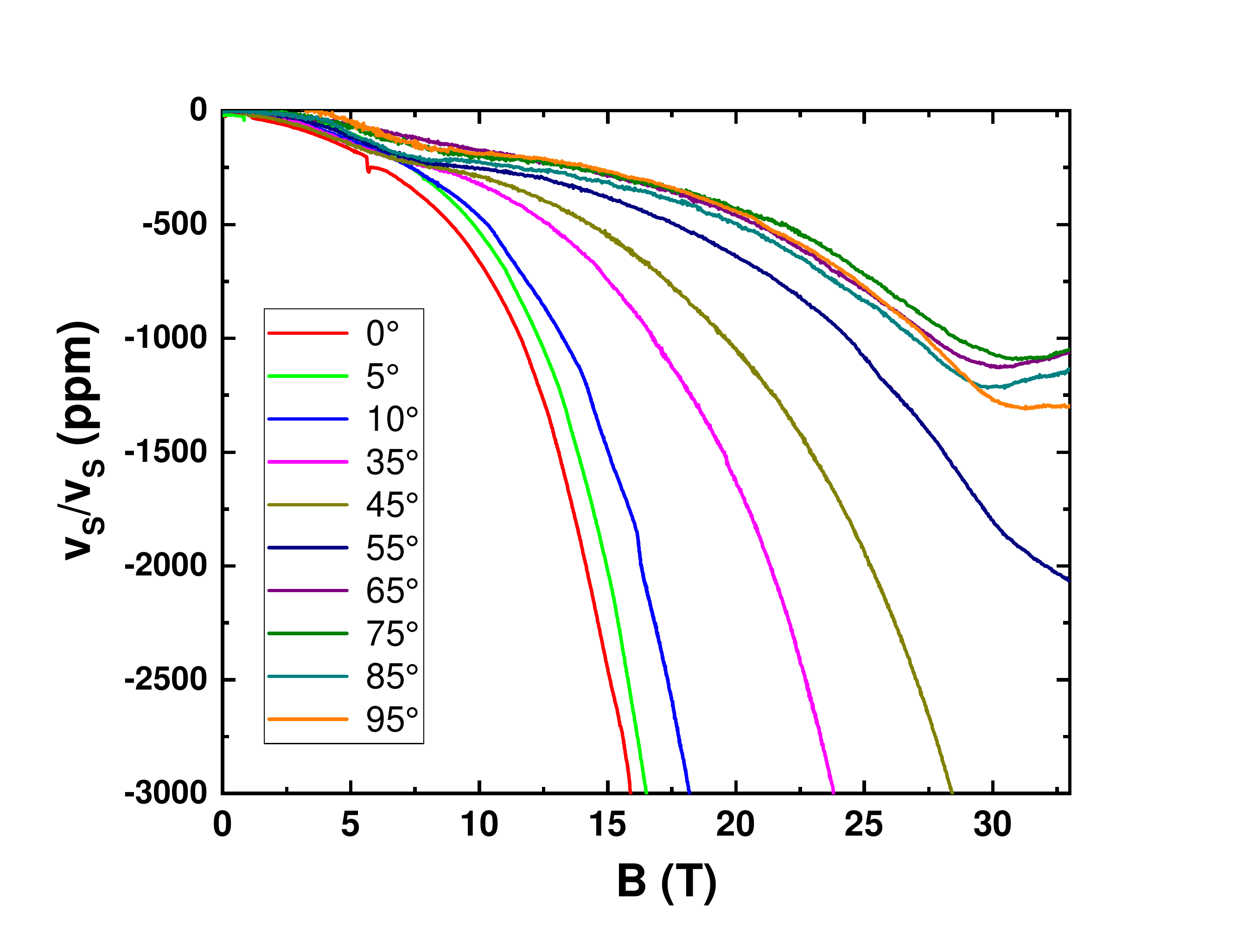}
\end{minipage}
\begin{minipage}[l]{0.4\textwidth}
\vspace{-0.5cm}
\hspace{-1cm}
\includegraphics[width=80mm]{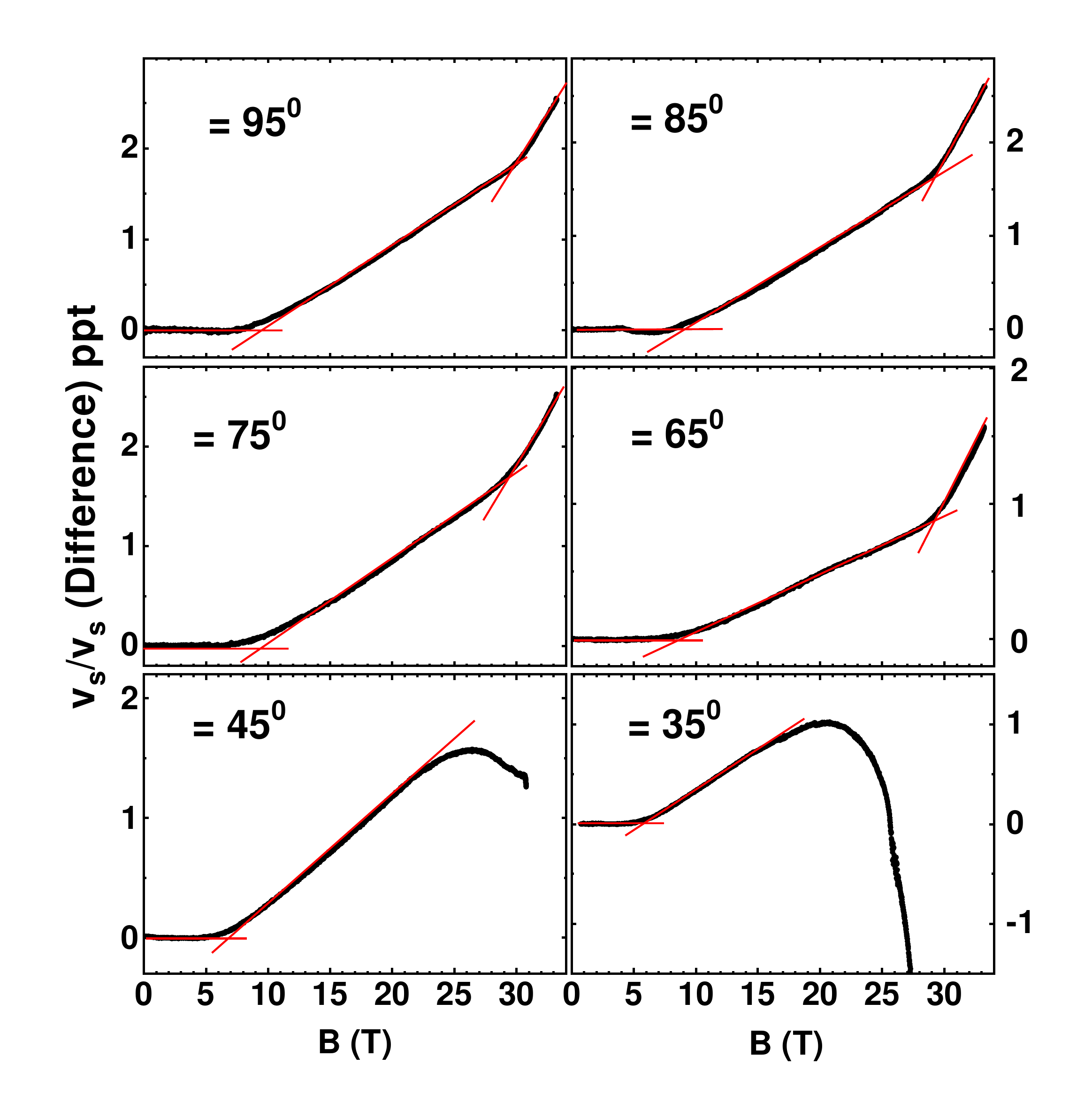}
\end{minipage}
\caption{\label{fig3}The top part of the figure shows sound velocity with q$\mid \mid$a measured for different angles of the applied magnetic field ranging from the a-axis (0$^\circ$) to the c-axis (90$^\circ$) at T = 0.5 K. For the a-axis the sound velocity shows a large decrease corresponding to the metamagnetic transition at 20 T. For increasing angle the curves shift to the right as the metamagnetic transition goes to higher field as 1/cos$\theta$. For angles larger than 45$^\circ$ instead of a continued steep downward trend in the velocity we see an upturn at larger fields.  For angles close to the c-axis features at $\approx$ 9T and 30T become apparent in the form of a dip similar to that shown in figure 1a. The bottom panel shows the data with a B$^2$ background subtracted similar to fig. S1.}
\end{figure}

\textbf{RESULTS:} In figure 1a we show the field dependence of the changes in the longitudinal sound velocity, $\delta v_s/v_s$, for 35 mK and 650mK.  Two dips in $ \delta v_s/v_s$ at $\approx$ 9T and 30T are apparent.  These begin at fairly high temperatures, get deeper as the temperature is lowered and the positions of these dips also change.  The maximum measured sound velocity changes are $\approx$ 200 ppm.  This means that it is important to obtain for any quantitative analysis of $\delta v_s/v_s$,  possible length changes in the sample which also contribute to the measured shifts. We do this through independently performed MS experiments.   The results for the sound velocity shown in fig.(1a) incorporate these length changes \cite{Note1}.  The MS data used to generate the corrected sound velocity changes are shown in fig.(1b).  The MS when the field and the length change are along the c-axis is always negative and follows the expected B$^2$ dependence.  The magnitude of the length changes decreases as the temperature is increased and is in agreement with previous work \cite{VisserJAP1987}. Also apparent in the figure at the lowest temperature shown, T=0.65 K, is a clear change in slope of MS from the initial B$^2$ behavior.  This departure sets in around 30 T, precisely where the higher field minimum in the sound velocity is observed. \\

It is tempting to construct a "phase diagram" from the observed features in the US and the MS.  Indeed, such an interpretation of the sound velocity minima was made by Bruls et al \cite{BrulsPhysicaB1996}.  However, caution is needed in such an interpretation since the position of the minima can be shifted by a "background" field dependent contribution \cite{AlferRubinJASA1954}.  Indeed, it is known that for both the US and MS this background occurs, due to trivial macroscopic effects, even in ordinary metals such as Cu and Ag, and is significantly different depending on the geometry i.e. whether the sound velocity (or the striction) is parallel or perpendicular to the field.  Therefore, a more appropriate procedure is to subtract the background contribution which for both the sound velocity and MS has a B$^2$ dependence.  Thus, least square fits to $\delta v_s/v_s$ were performed to the low field (i.e.  $\rm B <  3T$)  portion of the data to extract the macroscopic quadratic contribution.  Subtracting this, albeit extended to the entire measured field range, results in three separate linear regions which clearly define, via a change in slope between adjacent regions, the position of the observed features.  The result of such an analysis for the sound velocity is shown in the supplementary section, fig.S1.  The panel with the MS data at 0.65 K on the left in fig.1b illustrates similarly the three regions, demarkated by the two red arrows, seen after subtraction of a $\rm B^2$ contribution. The smaller feature (green arrow) corresponds to a Lifshitz transition which is discussed below.\\

The positions of the break in the linear regions of the sound velocity seen in fig.S1 are plotted in fig.1c.  Both the signatures referred to above move to lower fields as the temperature is increased and appear to originate at a temperature more closely linked to the Kondo scale in UPt$_3$ which is $\approx$ 20K. The 9 T feature as observed before\cite{BrulsPhysicaB1996} shifts to lower fields. Bruls et al have interpreted this low field feature as indicative of a transition from an SDW state which develops below the antiferromagnetic transition temperature of 5 K \cite{AeppliPRL1998, KoikeJPSJ1998}.  However, our measurements show this dip for B$\parallel$c persisting to well beyond 5 K and therefore appears unrelated to a possible long range AFM order in UPt$_3$. \\

On the other hand there is indeed a feature in our data that appears related to a possible 5K transition.  We note that the position of the high field feature takes a sharp upturn at  $\approx$4 K,  a temperature consistent with the SDW/AFM ordering in zero field.  Extrapolating the upper phase transition (prior to the upturn) to T=0  yields a magnetic field intercept of $\approx$24 T.  This field coincides with the small feature seen in the MS data referred to above and also marks the origin of very large amplitude magneto acoustic quantum oscillations (MAQO) as seen in the 35 mK scan shown in panel (a) of fig.1.  We show these oscillations more clearly in the top part of fig.2, after subtraction of a smooth background, as illustrated in fig.S2 \cite{Note2}.  The MAQO can be seen to suffer a frequency shift at 24.8 T in addition to the apparent large increase in the amplitude. Such a proximal occurance of pronounced quantum oscillations not necessarily coinciding with a phase transition is also seen in other HF systems close to a quantum critical point \cite{Jiao2015}.  These sudden changes are generally interpreted as Lifshitz transitions.  Other possibilities such as magnetic breakdown can be ruled out since our data on the high field side shows a very large amplitude.  Merging of Fermi surfaces due to magnetic breakdown would normally lead to a higher frequency with a lower amplitude instead \cite{Shoenberg1984}.  \\

Thus the abrupt increase in the magnitude of the MAQO can be explained in the context of a Lifshitz transition (LT) which implies a change in the topology of the Fermi surface (FS). To further interpret the MAQO and identify them with known orbits on the  FS of UPt$_3$ we performed a Fourier transform which can be seen in fig. 3. Clearly there are differences in the Fourier spectrum for the low field side B $<$ 24.8T and the high field side $\rm 24.8 T <B < 30 T$ of the LT.  There are two possiblities to assign the observed frequencies to the known orbits in $\rm UPt_3$.  According to Kimura et al. \cite{KimuraJPSJ1998} a hole orbit is present at 580 T corresponding to the "arm" of the octupus like FS of band 36.  On the other hand McMullan et al. \cite{McMullanNJP2008} find a similar frequency for the small electron "pearl"-like FS arising from band 39.  It is possible that a reconstruction of the FS at higher fields occurs such that the low frequency 240 T orbit vanishes.  The large amplitude 500 T oscillation in the region $\rm 24.8 T <B < 30 T$ may be identified with either the hole orbit or the electron orbit referred to above.  The sudden onset of the large amplitude could be the result of one or both physical effects:(i) it could arise from an abrupt reduction of the effective mass of the electrons on this part of the FS and (ii) since we are dealing with MAQO the readjusted FS may have an enhanced sensitivity to strain in the high field region.   Interestingly, the large oscillations dissappear again at 30T, the same magnetic field as the high field feature discussed above. Indeed, from the Fourier transform (fig. 2 bottom most panel) we again see a different frequency spectrum, suggesting the 30T feature is another LT.  As seen in fig.2 (top panel) all the LTs sharpen with decreasing temperature, as suggested by the orientation of the red arrows drawn parallel to the data, implying they are likely true quantum phase transitions.\\

We next turn to the behavior of the sound velocity for field directions away from the c-axis. The raw sound velocity data for various angles of the field are shown in fig.3-top panel.  The results obtained after removal of a quadratic background, similar to fig.S1, are shown in fig.3-bottom panel.  For small angles the upper signature remains more or less at the same value of 30 T.   However, for angles close to $\theta=51^0$ we lose the characteristic upturn (or hardening) in the sound velocity at the upper transition and instead the behavior is dominated by the steep down turn (or softening) which is the hallmark of the MM transition.  The transition points obtained from this angle dependent study are plotted in fig.4.  At the critical angle the two distinct transitions, the MM transition and the new 30 T transition come together marking a critical point.  At a critical angle of $\rm \approx 51^0$, the MM transition and the 30 T meet.  That the well known MM transition at 20 T for B$\parallel$ab-plane moves to higher fields with field angle tilt towards the c-axis was established by Suslov et al. \cite{SuslovIntJModPhys2002} through differential susceptibility measurements who verified the $\rm B_c$ proportional to 1/cos$\theta$ dependence.  In their work Suslov et al. also report a single data scan with ultrasound for $\rm \theta=60^0$ where a transition occurs at 37.6 T, shown as the ``star'' in fig.4. Thus the field-angle plane is demarkated into at least four regions.  The region between zero field and the low field transition which ranges from 4.5 T to 9 T as $\theta$ varies from $\rm 0^0$ to $\rm 90^0$ is the weak moment state identified in neutron scattering, which we label as $\rm SDW_A$.  This gives way to the region we label as $\rm SDW_B$ as the field inceases.  We note that there are no signatures in magnetization or any other thermodynamic quantity other than the sound velocity to mark this transition.  With a further increase in field $\rm SDW_B$ evolves into distinctly separate states depending on the angle of the magnetic field.  For $\rm \theta < 51^0$ via a discontinuous jump in the magnetization (MM transition) a polarized state, whose precise magnetic structure is yet to be measured in neutron diffraction, is reached.  Beyond  $\rm \theta = 51^0$ a new state which we label as $\rm SDW_C$ is established.  Since we observe the magnetization evolve smoothly across the transition to this new state, similar to the $\rm SDW_A$ to $ \rm SDW_B$ transition, our labelling is consistent. Apart from these four states it is also possible that there is a fifth state marked by the onset of the large MAQO and illustrated by the hatched region in fig.4.\\

\textbf{DISCUSSION:} Clues to the origin of such a unique critical point at an intermediate angle are provided by forced MS measurements carried out nearly three decades ago\cite{VisserJMMM1986}.  In their work deVisser et. al measured all four tensor components of the MS which we label as $\lambda_{a}^c$, $\lambda_{c}^c$, $\lambda_{a}^a$ and $\lambda_{c}^a$, where the subscripts refer to the direction of MS and the superscript refers to the magnetic field orientation.  Of the four components $\lambda_{a}^a$ and $\lambda_{a}^c$ are always positive i.e. no matter where the field points the basal plane always extends.  On the other hand, $\lambda_{c}^a$ is positive while $\lambda_{c}^c$ is negative i.e. extension/contraction along c depends on the field orientation.  This means that as one rotates B to some intermediate angle there should be a situation where the MS along c is precisely zero and all the field induced volume change comes from an extension in the plane.  Given the MS measurements of deVisser et al. this angle is calculated to be 51$^0$ away from the a-axis, precisely the angle where the three states come together to form the tricritical point (TCP) at 30 T. At this TCP, the $\rm SDW_B$-polarized state transition, the $\rm SDW_{B}-SDW_{C}$ transition, and the polarized state-$\rm SDW_C$ transition, all come together.  The $\rm SDW_B$-polarized state i.e. MM transition although perceived as a Kondo breakdown in earlier literature is more appropriately a LT involving a FS reconstruction \cite{Ludwig2019, BercxPRB2012} in $\rm UPt_3$.  Thus at the TCP three LT lines come together. \\

\begin{figure}
\includegraphics[width=0.5\textwidth]{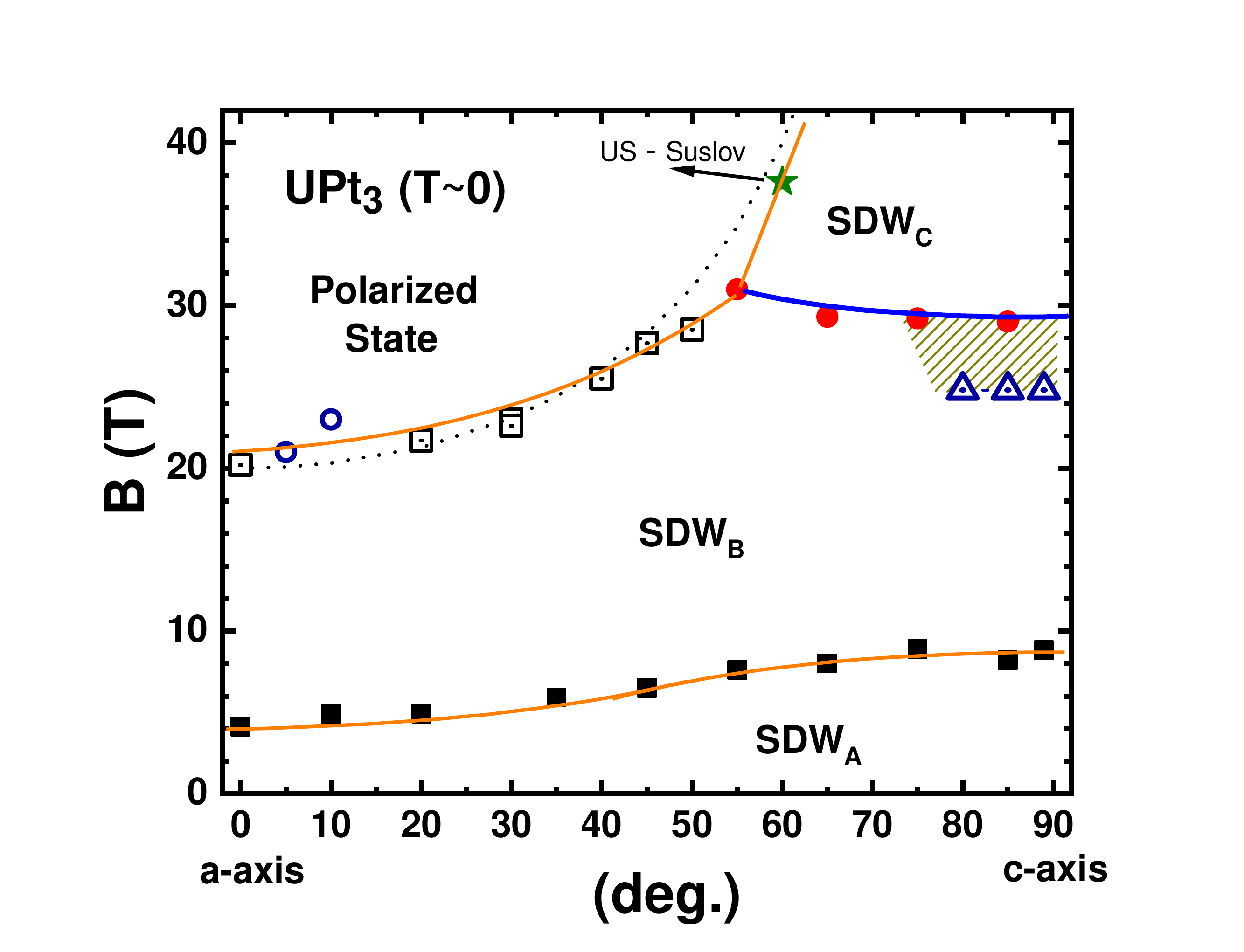}
\caption{\label{fig4} Phase diagram in the field - angle plane generated from a combination of magetometry and ultrasound velocity data.  The angle dependent magnetometry study by Suslov et al \cite{SuslovIntJModPhys2002} is used - open squares.  Although the magnetometry was confined to $\approx$ 1 K the position of the MM transition does not change considerably between 1K and 0.5 K, the latter being the temperature where the closed circles from US data were obtained. The single ``star'' at 37.6 T and $\rm \theta=60^0$ is from an ultrasound measurement by Suslov et al.  The closed squares, the open circles and the closed circles are from the current work.  The solid lines are guides to the eye demarkating the different states identified through the experimental points.  The dotted line represents the 1/cos$\theta$ dependence of the MM transition.  The inverted triangles mark the Lifshitz transition seen in US at 24.8 T. The hatched region represents another possible distinct state between the 24.8 LT and the 30 T transition.}
\end{figure}

Given the above experimental facts we can attempt to understand them in the context of existing theoretical ideas.  $\rm UPt_3$ traditionally has been regarded as a well behaved Fermi liquid system \cite{PethickPRL1986}.  However, there are many historical observations that are poorly understood.  The low moment magnetic state below 5 K presents no thumbprint in any thermodynamic measurement \cite{FisherSSC1991} or in ultrasound investigations where such transitions should generally be seen.  A so-called SDW transition however is seen in US at 4.5 T, $\rm \theta=0^0$, which shifts to 9 T for $\rm \theta=90^0$, but persists to T $>$ 5K, as discussed above.  Neutron scattering experiments performed in high fields up to 12 T, on the other hand, provide no clue about the nature of this SDW transition \cite{vanDijkPRB1998}.  Other experiments such as optical conductivity measurements indicate the presence of a psuedogap\cite{GrunerPRL1997} which develops roughly below 6 K in zero field.  In applied fields close to the MM transitition (B=20 T) a clear non-Fermi liquid behavior is observed in heat capacity measurements \cite{KimSSC2000}.  Interpreting these experimental facts can be challenging. Within a conventional approach considering spin fluctuations a theory for a quantum tricritical point has been proposed \cite{MisawaJPSJ2009}. There are specific predictions for the behavior of the inverse linear susceptibility, Hall coefficient, NMR relaxations times and heat capacity at the tricritical point that can be tested in $\rm UPt_3$.  Unconventional approaches \cite{SenthilPRB2004} with exotic excitations such as spinons have also been proposed to account for weak magnetism which could arise as a consequence of quantum fluctuations shielding the local moments. This approach as well as that based on local critical fluctuations \cite{Si2009} appear to provide the basis for a complex phase diagram with multiple spin states and subtle transitions between them.  Such subtlety is consistent with the experimental fact that the 9 T and the 30 T transitions present weak signatures only in US and MS measurements.  Extending such theoretical approaches to provide quantitiative predictions for thermodynamic and transport properties for the different states in the tri-critical region would be very useful for further comparison. \\
 
Apart from these aspects there are two other seemingly intertwined pieces of phenomenology revealed in the current experiments we wish to comment on.  First, the 3D to 2D crossover in magnetoelasticity occurring at $\theta=51^0$ due to the anisotropic response of magnetostriction is indeed a type of ``symmetry breaking''.  This is a special angle where the notion of Poisson's effect apparently breaks down and magnetoelastic effects are likely to be dominated by two dimensional magnetization (spin) fluctuations. Second, this ``symmetry breaking'' is electronically driven and does not have a structural origin.  We can state that since FS oscillation frequencies measured at all angles are by and large in agreement with topology from band structure calculations performed assuming a hexagonal crystal structure \cite{KimuraJPSJ1998}.  Clearly this is an effect that need not be unique to $\rm UPt_3$. It is also interesting to note that the elastic softening at the tricritical point observed in the present work is reminiscent of the breakdown of Hooke's law due to the electronic Mott transition, an isostructural solid-solid transition -  also the case here, observed in an organic conductor under pressure recently \cite{GatiSciAdv2016, ZachariasEPJ2015}.  \\

In conclusion, through new high resolution measurements of the ultrasound velocity, magnetostriction and magnetization, we have established several new spin states in high magnetic fields at varying angles in the prototypical strongly correlated metal $\rm UPt_3$.  We have also identified a unique quantum critical point which arises at an orientation of the magentic field intermediate between the c-axis and the basal plane thus ignoring the underlying crystal symmetry.  At this critical point an apparent 3D-2D magnetoelastic crossover occurs and a large elastic instability arises.  Further studies of this critical elasticity and the associated FS instability are necessary for a more complete understanding of the different spin states in $\rm UPt_3$.  Along with high field neutron scattering experiments future work should focus on measurements of the Hall response, the magnetic susceptibility, dHVA effect and ultrasound attenuation. Such a synthetic approach is mandatory for a satisfactory understanding of the complex states and phenomenology in strongly correlated quantum materials such as $\rm UPt_3$.\\

\textbf{METHODS}: The single crystals of UPt$_3$ were obtained through float zone refinement of a polycrystalline rod cast in an arc melter.  The ultrasound velocity data was obtained by employing a frequency modulated continuous wave ultrasonic technique where shifts in the standing wave resonance are measured in a $\rm UPt_3$ crystal of roughly 3 mm x 3mm x 3mm size.  The measurements were performed at operating ultrasound frequencies between 19 MHz and 61 MHz.  The MS measurements were carried out at the Los Alamos pulsed field facility in fields up to 65 T and temperatures down to 0.58 K using a fiber Bragg grating interferometric method \cite{Jaime2017}. \\

\textbf{Data Availability} Data will be made available on reasonable request through email to the lead author (BSS).

\textbf{ACKNOWLEDGEMENTS} We thank David Hinks for the crystal of $\rm UPt_3$ and Eric Palm and Tim Murphy in Tallahassee and John Betts at Los Alamos for assisting with experiments.  We are particularly grateful to Andrey Chubukov, Qimiao Si, Anne de Visser, Vittorio Celli, and Gia-Wei Chern for many useful conversations.  The National Magnet Laboratory in Tallahassee and at Los Alamos is supported by the National Science Foundation and the State of Florida. The work at the University of Virginia was funded by NSF DMR 0073456.\\

\end{document}



\begin{center}
\textbf{Supplementary Information: Field Angle Tuned Metamagnetism and Lifshitz Transitions in UPt$_3$}\\
{B.S. Shivaram$^1$, Ludwig Holleis$^1$, V.W. Ulrich$^1$, Marcelo Jaime$^2$ and John Singleton$^2$}\\
{$^1$Department of Physics, University of Virginia, Charlottesville, VA. 22904}\\
{$^2$National High Magnetic Field Laboratory, Los Alamos National Labs, Los Alamos, New Mexico}\\

\end{center}

\textbf{Ultrasound Velocity - subtraction of quadratic field dependence} \\ 
As stated in the main text, for the curves shown in fig.1a (top left panel) of the main text, which shows the change in the sound velocity, we perform a parabolic fit to only the low field portion (B$<3$ T).  The obtained quadratic field dependence is subtracted from each curve to produce the results shown below, fig.S1.  Post subtraction the sound velocity change falls into three distinct linear regions. The transition points between these regions as defined by the intersection of the blue lines are used to construct the B-T phase diagram shown in fig.1c.   \\

\begin{figure}[h]
\includegraphics[width=180mm]{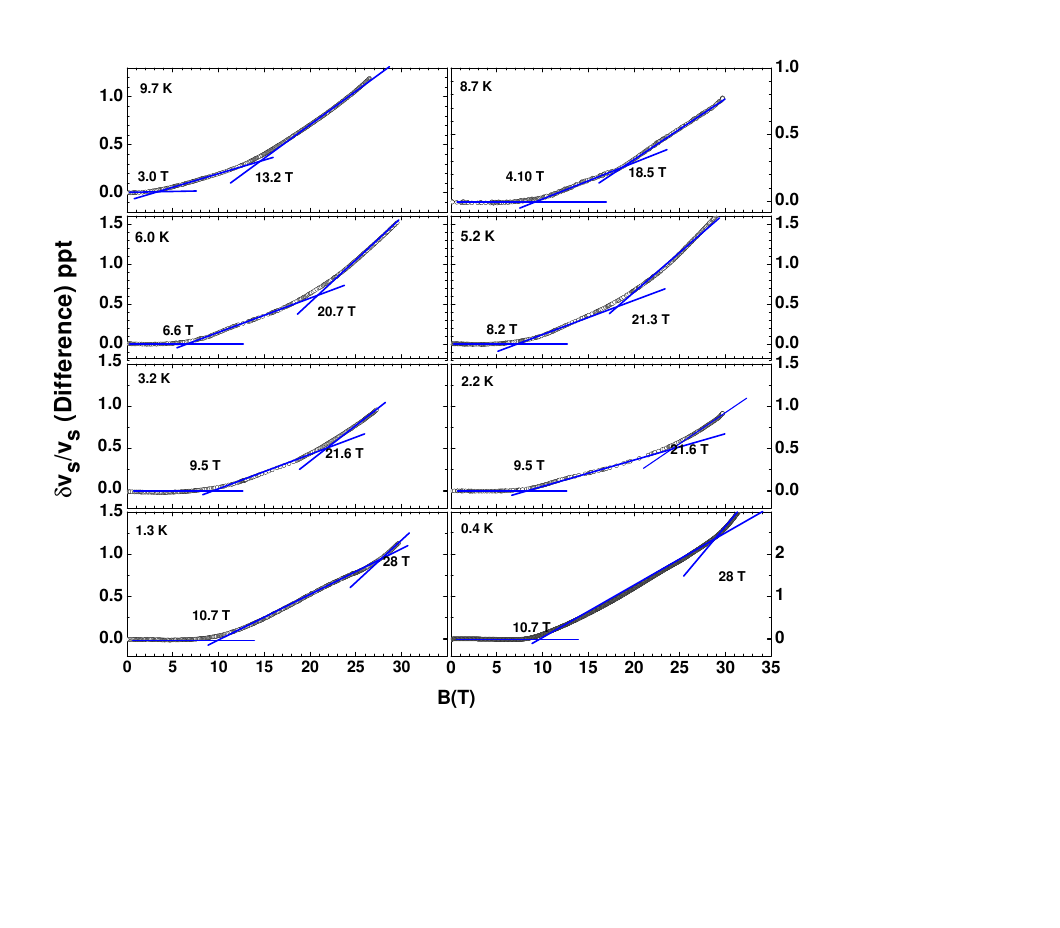}
\setlength{\abovecaptionskip}{-90pt}
\caption{Shows the sound velocity replotted after the subtraction of an initial background B$^2$ dependence.  Note that the sound velocity increases post both transitions.}
\end{figure}

\textbf{Magneto Acoustic Quantum Oscillations - Background subtraction}\\
In the four different panels in the figure below we illustrate step-by-step the procedure we followed to subtract the background prior to performing a Fourier transformation of the MAQO data.

\begin{figure}[h]
\includegraphics[width=180mm]{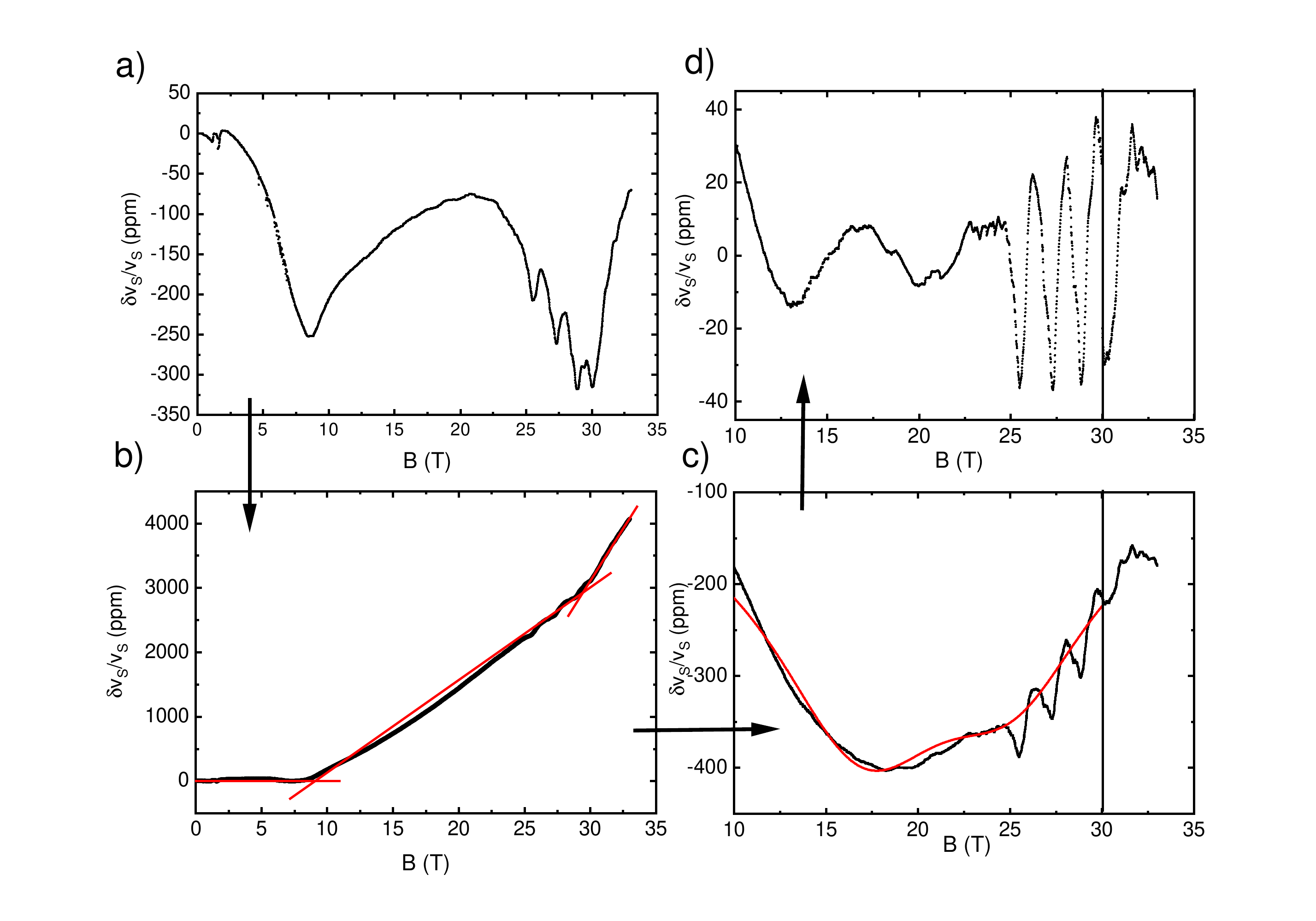}
\setlength{\abovecaptionskip}{0pt}
\caption{Panel shows the as obtained data on the sound velocity at T=35 mK.  A subtraction of an initial background B$^2$ dependence results in the plot shown in panel (b).  Subtraction of a linear fit to the intermediate region, 10 T to 30 T and high field region 30 T-33 T, results in a remanent part shown in panel (c).  A smoothly varying double Lorentzian (inverted) is further subtracted to obtain the results shown in panel (d) on which the Fourier transformation is carried out.}
\end{figure}